\documentclass{pnastwobis}
\pdfoutput=1

\usepackage{amssymb,amsfonts,amsmath}
\usepackage[pdftex]{graphicx}
\usepackage{multirow}
\usepackage{color}

\begin{document}

\title{Extracting the multiscale backbone of complex weighted networks}

\author{M. \'Angeles Serrano
\thanks{To whom correspondence should be addressed. E-mail: marian.serrano@ifisc.uib-csic.es}
\affil{1}{IFISC Instituto de F\'{i}sica Interdisciplinar y Sistemas Complejos
(CSIC-UIB), Campus Universitat Illes Balears, E-07122 Palma de Mallorca, Spain},
Mari{\'a}n Bogu{\~n}{\'a} \affil{2}{Departament de F{\'{\i}}sica Fonamental,
Universitat de Barcelona, Mart\'{\i} i Franqu\`es 1, 08028 Barcelona, Spain}, \and
Alessandro Vespignani \affil{3}{Center for Complex Networks and Systems Research,
School of Informatics, Indiana University, 919 E. 10th Street, Bloomington, IN
47406, USA} \affil{4}{Complex Networks Lagrange Laboratory (CNLL), Institute for
Scientific Interchange (ISI), Torino, Italy}}

\maketitle

\begin{article}

\begin{abstract}
A large number of complex systems find a natural abstraction in the form of
weighted networks whose nodes represent the elements of the system and the weighted
edges identify the presence of an interaction and its relative strength. In recent
years, the study of an increasing number of large scale networks has highlighted
the statistical heterogeneity of their interaction pattern, with degree and weight
distributions which vary over many orders of magnitude. These features, along with
the large number of elements and links, make the extraction of the truly relevant
connections forming the network's backbone a very challenging problem. More
specifically, coarse-graining approaches and filtering techniques are at struggle
with the multiscale nature of large scale systems. Here we define a filtering
method that offers a practical procedure to extract the relevant connection
backbone in complex multiscale networks, preserving the edges that represent
statistical significant deviations with respect to a null model for the local
assignment of weights to edges. An important aspect of the method is that it does
not belittle small-scale interactions and operates at all scales defined by the
weight distribution. We apply our method to real world network instances and
compare the obtained results with alternative backbone extraction techniques.
\end{abstract}

\keywords{disordered systems | multiscale phenomena | filtering
  |visualization }

\dropcap{I}n recent years, a huge amount of data on large scale social, biological,
and communication networks, meticulously collected and catalogued, has become
available for scientific analysis and study. Examples can be found in all domains;
from technological to social systems and  transportation networks on a local and
global scale, and down to the microscopic scale of biochemical
networks~\cite{Newman:2003,Dorogovtsev:2007a,Caldarelli:2007}. Common traits of
these networks can be found in the statistical properties characterized by large
scale heterogeneity with statistical observables such as nodes' degree and traffic
varying over a wide range of scales~\cite{ Barabasi:1999}. The sheer size and
multiscale nature of these networks make very difficult the extraction of the
relevant information that would allow a reduced representation while preserving the
key features we want to highlight. A typical example is faced in the visualization
of networks. While it is generally possible to create wonderful images of large
scale heterogeneous networks, the amount of valuable information gathered is in
most cases very little because of the redundant intricacy generated by the
overwhelming number of connections. Problems such as the extraction of the relevant
backbone or the isolation of the statistically relevant structures/signal that
would allow reduced but meaningful representations of the system are indeed major
challenges in the analysis of large-scale networks.

In complex weighted networks, the discrimination of the right trade-off between the
level of network reduction and the amount of relevant information preserved in the
new representation faces us with additional problems. In many cases, the
probability distribution $P(\omega)$ that any given link is carrying a weight
$\omega$ is broadly distributed, spanning several orders of magnitude. This feature
implies the lack of a characteristic scale and any method based on thresholding
would simply overlook the information present above or below the arbitrary cut-off
scale. While this issue would not be a major drawback in networks where the
intensities of all the edges are independently and identically distributed, the cut
off of the $P(\omega)$ tail would destroy the multiscale nature of more realistic
networks where weights are locally correlated on edges incident to the same node
and non-trivially coupled to topology~\cite{Barrat:2004b}. Thus, the presence of
multiscale fluctuations calls for reduction techniques that consistently highlight
the relevant structures and hierarchies without favoring any particular resolution
scale. Furthermore, it also demands to change the focus towards a local perspective
rather than a global one, where the relevance of the connections could be decided
at the level of nodes in relative terms.

In this work, we concentrate on a particular technique that operates at all the
scales defined by the weighted network structure. This method, based on the local
identification of the statistically relevant weight heterogeneities, is able to
filter out the backbone of dominant connections in weighted networks with strong
disorder, preserving structural properties and hierarchies at all scales. We
discuss our multiscale filter in relation to the appropriate null model that
provides the basis for the statistical significance of the heterogeneity
measurements. We apply the technique to two real world networks, the U.S. airport
network and the Florida Bay food web, and compare the results to those obtained by
the application of thresholding methods.

\section{Results and Discussion}
In Statistical Mathematics, as in other areas, filtering techniques aimed at
uncovering the relevant information in data sets are popular and successful. One
could cite, for instance, the Principal Components Analysis to identify hidden
patterns by reducing the effective dimension of multivariate
data~\cite{Jolliffe:2002}. In the following, we will refer to the {\em network
reduction} as the construction of a network that contains far less data (in our
case links) and allows the discrimination and computational tractability of the
relevant features of the original networks; for instance, the traffic backbone of a
large scale transportation infrastructure. Reduction schemes can be divided into
two main categories: coarse-graining and filtering/pruning. In the first case,
nodes sharing a common attribute could be gathered together in the same class
--group, community, etc.-- and then substituted by a single new unit which
represents the whole class in a new network representation of the
system~\cite{Kim:2004,Song:2005,Itzkovitz:2005,Gfeller:2007}. This coarse-graining
is indeed zooming out the system so that it can be observed at different scales.
Something completely different is done when a filter is applied. In this case, the
observation scale is fixed and the representation that the network symbolizes is
not changed. Instead, those elements --nodes and edges-- that carry relevant
information about the network structure are kept while the rest are discarded. An
example of a well-known hierarchical topological filter, although usually not
referred as such, is the $k$-core decomposition of a network~\cite{Chalupa:1979},
with a filtering rule that acts on the connectivity of the nodes.

In the case of weighted networks~\cite{Barrat:2004b}, two basic reduction
techniques refer to the extraction of the minimum spanning tree and the application
of a global threshold on the weights of the links so that just those that beat the
threshold are preserved. The minimum spanning tree of a graph $\cal{G}$, a
classical concept of graph theory~\cite{Kruskal:1956}, is the shortest length tree
subgraph that contains all the nodes of $\cal{G}$. These definitions can be
generalized for weighted graphs~\cite{Macdonald:2005}. A minimum spanning tree of a
weighted graph $\cal{G}$ is the spanning tree of $\cal{G}$ whose edges sum to
minimum weight. This idea has been exploited along with percolation criticality to
define superhighways in weighted networks~\cite{Wu:2006}. By using opportune
transformation rules for the weights, it is also possible to define maximum
weighted spanning trees and other analogous definitions. One of the big limitations
of this method is that spanning trees are by construction acyclic. This means that
reduced networks obtained by this algorithm are overly structural simplifications
that destroy local cycles, clustering coefficient and the clustering hierarchies
often present in real world networks.

These previous drawbacks are not present in the application of a threshold to the
global weight distribution that removes all connections with a weight below a given
value $\omega_{c}$. This filter has been used for instance in the study of
functional networks connecting correlated human brain sites~\cite{Eguiluz:2005} and
food web resistance as a function of link magnitude~\cite{Allesina:2006}. This
approach, however, belittles nodes with a small strength $s$ (defined as the sum of
weights incident to the node $s_i=\sum_j w_{ij}$), since the introduction of
$\omega_{c}$ induces a characteristic scale from the outset. As a consequence,
strongly disordered networks with heavy-tailed statistical distributions $P(s)$ and
$P(\omega)$ make this simple thresholding algorithm very poorly performing since
nodes with small $s$ are systematically overlooked. This is even a more serious
drawback when weights are correlated at the local level. In this type of networks,
interesting features and structures are present at all scales and the introduction
of such artificial cut-off drastically removes all information below the cut-off
scale.

\subsection{Local fluctuations} In order to develop a multiscale reduction
algorithm, we take advantage of the local fluctuations of weights on the links
emanated by single nodes. In heterogeneous weighted networks with strong disorder,
i.e. heavy tailed $P(\omega)$ and $P(s)$ distributions, a few links carry the
largest proportion of the node's total strength. Furthermore, most real networks
have nodes surrounded by incident edges with associated weights that are
heterogeneously distributed and correlated between them. The fingerprint of these
correlations is observed in the non-trivial dependence between weights and
topology~\cite{Barrat:2004b}. The better a node is connected to the rest of the
network, the higher the weight of its edges so that the strength tends to grow
superlinearly with the degree. However, the strength alone is not enough to capture
the weighted structure of nodes even at the local level. We need to introduce some
measure of the fluctuations of the weights attached to a given node, and we want to
do it at the local level in relative terms so that each node could independently
assess the importance of its connections. To this end, we first normalize the
weights of edges linking node $i$ with its neighbors as $p_{ij}=\omega_{ij}/s_i$,
being $s_i$ the strength of node $i$ and $w_{ij}$ the weight of its connections to
its neighbor $j$. Then, by using the disparity function defined in the Materials
and Methods section, it is possible to see that even at the local level defined by
the edges adjacent to a single node a few of those edges carry a disproportionate
fraction $p_{ij}$ of the node's strength, with the remaining edges carrying just a
small fraction of the node's strength~\cite{Guichard:2003,Almaas:2004}.

Being more specific, we are interested in all edges with weights representing a
significant fraction of the local strength and weight magnitude of each given node.
However, local heterogeneities could simply be produced by random fluctuations. It
is then fundamental to introduce a null model that informs us about the random
expectation for the distribution of weights associated to the connections of a
particular node. Empirical values not statistically compatible with the null model
define, on a node by node basis, whether the observed weight heterogeneity and
intensity are statistically significant and define the relevant part of the signal
due to specific and relevant organizing principles of the network structure. This
procedure would determine without arbitrariness how many connections for every node
belong to the backbone of connections that carry a statistically disproportionate
weight --be them one, zero or many--, providing sparse subnetworks of connected
links selected according to the total amount of weight we intend to characterize.
This reduction scheme necessarily encodes a wealth of information as the reduced
network does not contain only the links carrying the largest weight in the network
but also all links which can be considered, according to a pre-defined statistical
significance level, defining the  relevant structure (signal) generated by the
weight and strength assignment with respect to the simple randomness of the null
hypothesis. An important aspect of this construction is that the ensuing reduction
algorithm does not belittle small nodes in terms of strength and then offers a
practical procedure to reduce the number of connections taking into account all the
scales present in the system.

\subsection{The disparity filter} In the following, we
discuss the disparity filter for undirected weighted networks, although it is also
applicable to directed ones as reported in the Supporting Information. The null
model that we use to define anomalous fluctuations provides the expectation for the
disparity measure of a given node in a pure random case. It is based on the
following null hypothesis: the normalized weights which correspond to the
connections of a certain node of degree $k$ are produced by a random assignment
from a uniform distribution. To visualize this process, $k-1$ points are
distributed with uniform probability in the interval $[0,1]$ so that it ends up
divided in $k$ subintervals. Their lengths would represent the expected values for
the $k$ normalized weights $p_{ij}$ according to the null hypothesis. The
probability density function for one of these variables taking a particular value
$x$ is
\begin{equation}
\rho(x)dx=(k-1)(1-x)^{k-2}dx, \label{eq:PX}
\end{equation}
which depends on the degree $k$ of the node under consideration. In the Material
and Methods section we provide a detailed analysis of the null model with respect
to the actual weight distribution in two real world networks.

The disparity filter proceeds by identifying which links for each node should be
preserved in the network. The null model allows this discrimination by the
calculation for each edge of a given node of the probability $\alpha_{ij}$ that its
normalized weight $p_{ij}$ is compatible with the null hypothesis. In statistical
inference, this concept is known as the $p$-value, the probability that, if the
null hypothesis is true, one obtains a value for the variable under consideration
larger or equal than the observed one. By imposing a significance level $\alpha$,
the links that carry weights which can be considered not compatible with a random
distribution can be filtered out with an certain statistical significance. All the
links with $\alpha_{ij}<\alpha$ reject the null hypothesis and can be considered as
significant heterogeneities due to the network organizing principles. By changing
the significance level we can filter out the links progressively focusing on more
relevant edges. The statistically relevant edges will be those whose weight satisfy
the relation
\begin{equation}
\alpha_{ij}=1-(k-1)\int_0^{p_{ij}} (1-x)^{k-2}dx < \alpha.
\label{eq:confidencelevel}
\end{equation}
Note that this expression depends on the number of connections $k$ of the node to
which the link under consideration is attached.

The multi-scale backbone is then obtained by preserving all the links which satisfy
the above criterion for at least one of the two nodes at the ends of the link while
discounting the rest~\footnote{In the case of a node $i$ of degree $k_{i}=1$
connected to a node $j$ of degree $k_{j}>1$, we keep the connection only if it
beats the threshold for node $j$}. In this way, small nodes in terms of strength
are not belittled so that the system remains in the percolated phase. In other
words, we single out the relevant part of the network that carries the
statistically relevant signal provided by the distribution with respect to a local
uniform randomness null hypotheses. By choosing a constant significance level
$\alpha$ we obtain a homogeneous criterion that allows us to compare
inhomogeneities in nodes with different magnitude in degree and strength.
Decreasing the statistical confidence more restrictive subsets are obtained, giving
place to a potential hierarchy of backbones. This strategy will be efficient
whenever the level of heterogeneity is high and weights are locally correlated.
Otherwise, the pruning could lose its hierarchical attribute producing analogous
results to the global threshold algorithm (see section ``Networks with uncorrelated
weights'' in Supporting Information).

\subsection{The multiscale backbone of real networks} To test the
performance of the disparity filter algorithm, we apply it to the extraction of the
multiscale backbone of two real world networks. We also compare the obtained
results with the reduced networks obtained by applying a simple global threshold
strategy that preserves connections above a given weight $\omega_{c}$. As examples
of strongly disordered networks, we consider the domestic non-stop segment of the
U.S. airport transportation system for the year 2006~\cite{USANData} and the
Florida Bay ecosystem in the dry season~\cite{Ulanowicz:1998}. The U.S. airport
transportation system for the year 2006 gathers the data reported by air carriers
about flights between $1078$ USA airports connected by $11890$ links. Weights are
given by the number of passengers traveling the corresponding route in the year
symmetrized to produce an undirected representation. The resulting graph has a high
density of connections, $\langle k \rangle =22$, making difficult both its analysis
and visualization. The Florida Bay foodweb comes from the ATLSS Project by the
University of Maryland~\cite{FWData}. Trophic interactions in food webs are
symbolized by directed and weighted links representing carbon flows
($mgCy^{-1}m^{-2}$) between species. The network consists of a total of 122
separate components joined by 1799 directed links.
\begin{table}[t]
\caption{Sizes of the disparity backbones in terms of the percentage of total
  weight ($\%W_T$),
 nodes ($\%N_T$), and edges ($\%E_T$) for different values of the significance
 level $\alpha$. See points (a) and (b) in Fig.3.}
\vspace{-0.2cm}
\begin{center}
\begin{tabular}{lcccclccc}
\hline \hline\\[-0.99mm]
\multicolumn{4}{c}{US Airport Network}&&\multicolumn{4}{c}{Florida Bay Food Web}\\
\cline{1-4}\cline{6-9}\\[-0.99mm]
$\alpha$&\scriptsize $\%W_T$& \scriptsize$\%N_T$& \scriptsize$\%E_T$
&&$\alpha$&\scriptsize$\%W_T$& \scriptsize$\%N_T$& \scriptsize$\%E_T$\\
\hline \hline
0.2&94& 77& 24&&0.2&90& 98& 31\\
0.1&89& 71& 20&&0.1&78& 98& 23\\
0.05(a)&83& 66& 17&&0.05&72& 97& 16\\
0.01&65& 59& 12&&0.01&55& 87& 9\\
0.005& 58& 56&10&&0.0008(a)&49& 64& 5\\
0.003(b)& 51& 54&9&&0.0002(b)&43& 57& 4\\
\hline \hline
\end{tabular}
\end{center}
\label{table_conf}
\end{table}

In Table~1 and Fig.~1, we show statistics for the relative sizes --in terms of
fractions of total weight $W_{T}$, nodes $N_{T}$, and edges $E_{T}$-- preserved in
the backbones when the network is filtered by the disparity filter and by the
application of a global threshold, respectively. The disparity filter reduces the
number of edges significantly even when the significance level $\alpha$ is close to
1, keeping at the same time almost all the weight and a high fraction of nodes.
Smaller values of $\alpha$ reduce even more the number of edges but, interestingly,
the total weight and number of nodes remain nearly constant. Only for very low
values of $\alpha$ --when the filter becomes very restrictive-- the total weight
and number of nodes start decreasing significantly. In the case of the airports
network, values around $\alpha \approx 0.05$ extract backbones with more than
$80\%$ of the total weight, $66\%$ of nodes, and only $17\%$ of edges. The global
threshold filter, on the other hand, is not able to maintain the majority of the
nodes in the backbone for similar values of retained weight or edges, as it is
clearly seen in the first and second columns of Fig.~1, respectively.

It is particularly interesting to analyze the behavior of the topological
properties of the filtered network at increasing levels of reduction. Fig.~2 shows
the evolution of the cumulative degree distribution, {\it i. e.}
$P_c(k)=\sum_{k'\ge k} P(k')$, for different values of $\alpha$ (left top plot) and
$\omega_{c}$ (right top plot), respectively. The original airports network is heavy
tailed although cannot be fitted by a pure power law function. Interestingly, the
disparity filter reveals a clear power law behavior as $\alpha$ decreases, with an
exponent $\gamma \approx 2.3$. On the other hand, the global threshold filter
produces subgraphs with a degree distribution similar to the original one but with
a sharp cut-off that becomes smaller as the filter gets more restrictive. On the
other side, the weight distribution $P(\omega)$ for the disparity filter (left
middle plot) shows that almost all scales are kept during the filtering process and
only the region of very small weights is affected, in contrast to the global
threshold filter that, by definition, cuts $P(\omega)$ off below $\omega_{c}$
(middle right plot).

In the bottom plots of Fig.~2, we show the clustering coefficient $C$ measured as
the average over nodes of degree larger than $1$. It remains nearly constant in
both filters until they become too restrictive, in which case clustering goes to
zero\footnote{The sudden increase of clustering for $E_{B}/E_{T}=0.2$
  is due to the reduction of the number of nodes in the network,
  increasing then the chances of having a random contribution.}. In
the case of the disparity filter, clustering remains constant up to values of
$\alpha \approx 0.01$. This is precisely the value below which both the number of
nodes and the weight in the backbone start decreasing significantly. Therefore, we
can conclude that values of $\alpha$ in the range $[0.01,0.5]$ are optimal, in the
sense that backbones in this region have a large proportion of nodes and weight,
the same clustering of the original network, and a stable stationary degree
distribution, all with a very small number of connections as compared to the
original network. It is important to stress that the disparity filtering also
includes the connections with the largest weight present in the system. This is
because the heavy-tail of the $P(\omega)$ distribution is mainly determined by
relevant large-scale weight. This is clearly illustrated in Fig.~3, where we show
that for statistical significance levels up to $\alpha\simeq 10^{-3}$, all the
edges included in the 10-20\% of the $P(\omega)$ tail are included in the extracted
multiscale backbone.
\begin{figure}[t]
\begin{center}
\includegraphics[height=7cm]{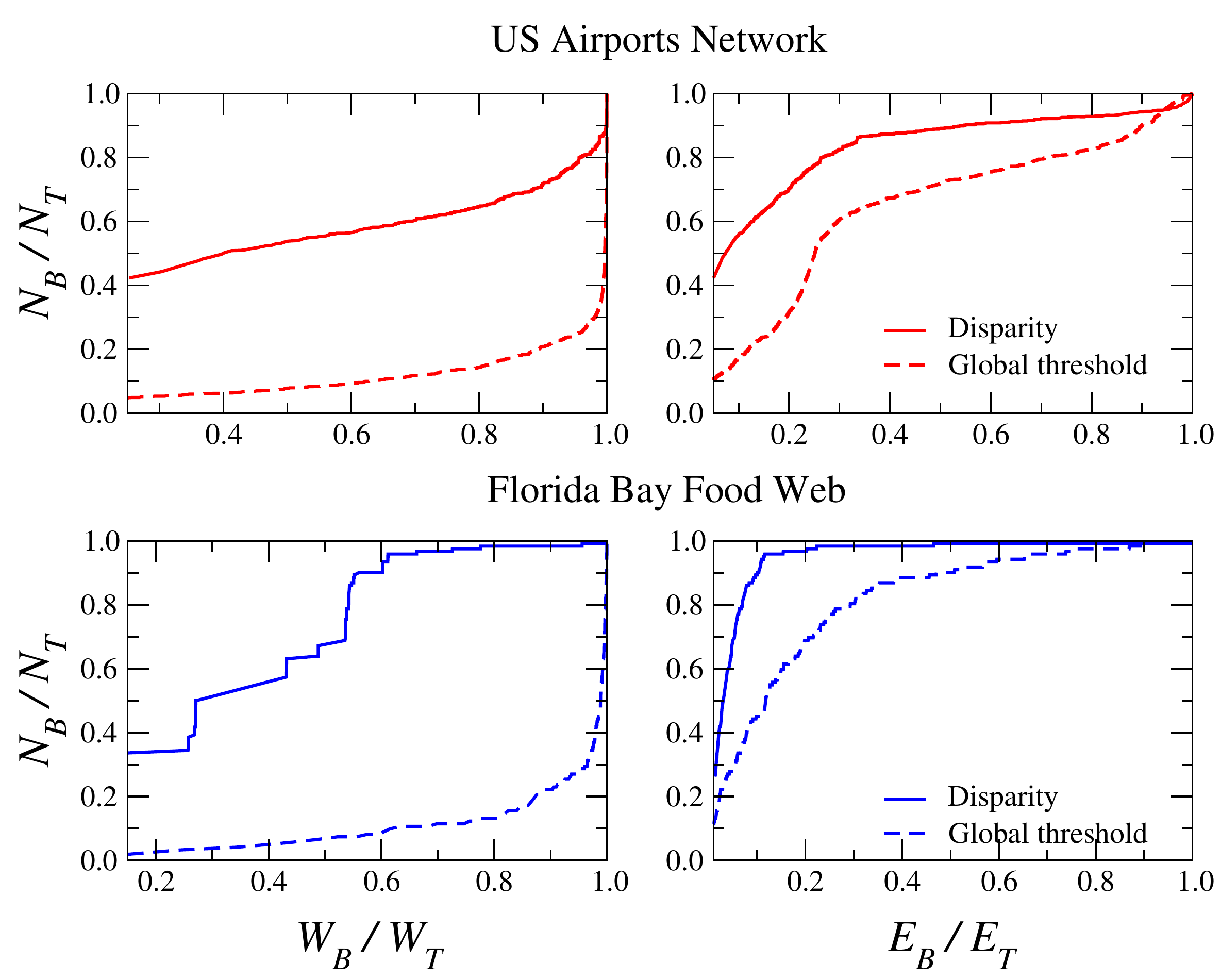}
\end{center}
\caption{Fraction of nodes kept in the backbones as a function of the
  fraction of weight (left) and edges (right) retained by the
  filters.}
\label{fig:1}
\end{figure}

As an illustration of the efficacy of the disparity filter, we visualize the
obtained multi-scale backbone in Fig.~4. In the case of the US airport network we
use the significance value $\alpha=0.003$ (see entry (b) in Table~1 and Fig.~3).
Interestingly, the disparity filter offers a perspective of the network that
reveals its geographic constrains (notice that each node is placed in the plane
according to its actual coordinates on the earth). It is possible to identify local
hubs with very well defined basins of attraction made of small airports connected
to them~\cite{Barthelemy:2006}, a star-like pattern that is particularly clear in
Alaska airports or mid west cities. In addition, the hierarchy of the
transportation system is fully highlighted, including not just the most high flux
connections but also small weight edges which are statistically significant as they
represent relevant signal at the small scales. In this way, all important
connection on the local and global level are considered at once. This would not be
possible with a global threshold algorithm, that would simply eliminate all
connections below the scale introduced by the cut-off threshold.
\begin{figure}[t]
\begin{center}
\includegraphics[height=11cm]{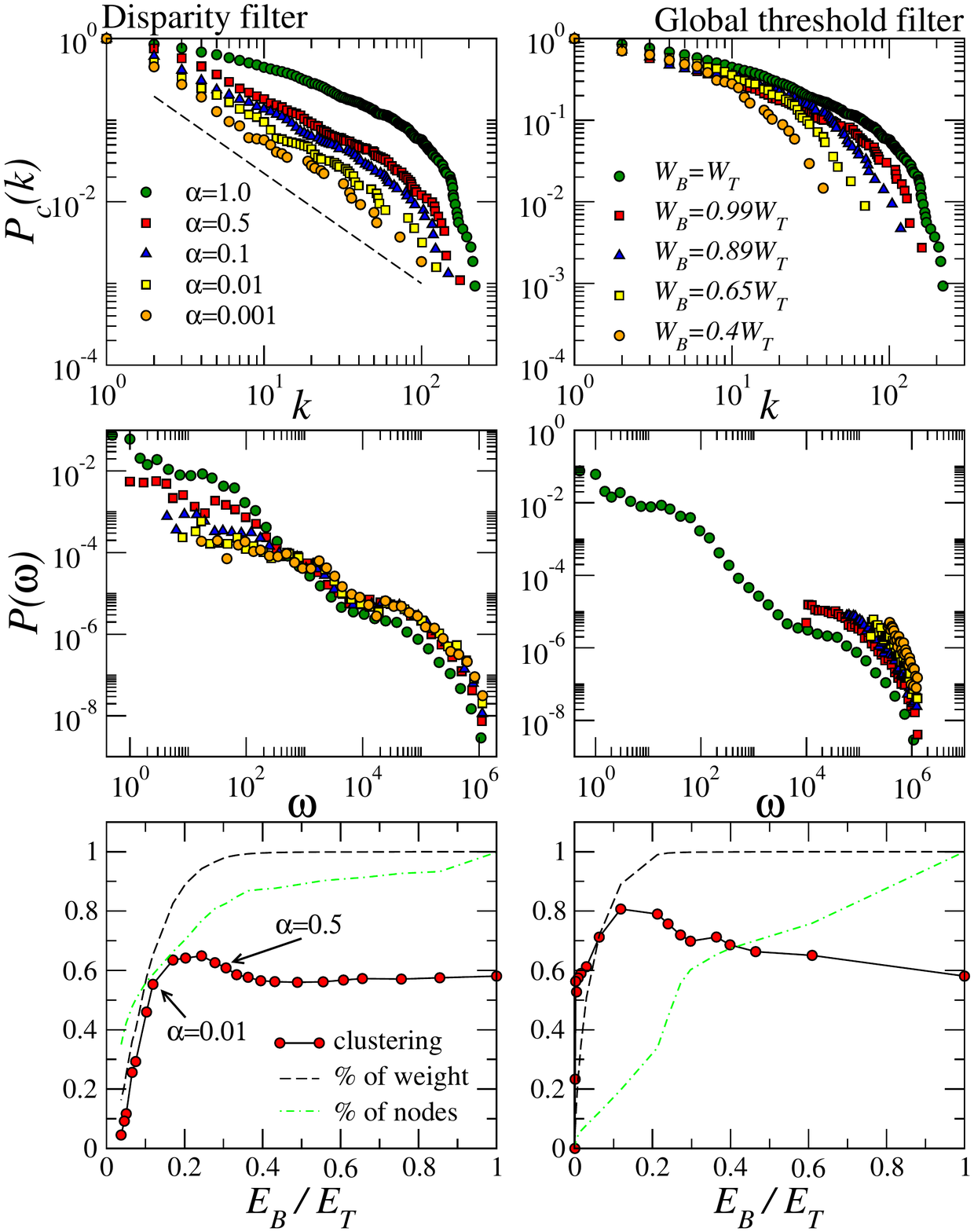}
\end{center}
\caption{Topology of the filtered subgraphs for the U.S. airports
  Network. {\bf Top:} Cumulative degree distribution, $P_c(k)$, for
  the disparity (left) and global threshold (right) backbones. The
  values of $\omega_c$ on the right plot are chosen to generate
  subgraphs with the same weight as the ones shown on the left
  plot. {\bf Middle.} Distribution of links' weights of the different subgraphs generated by the two filters. Symbols are the same as in the top plots.  {\bf Bottom.} Clustering coefficient averaged over nodes of
  degree larger than 1 for the two methods as a function of the fraction of edges in the backbones. Dashed lines show the fraction of nodes and weight for a given fraction of edges.} \label{topology_filtered}
\end{figure}

\begin{figure}[t]
\begin{center}
\includegraphics[height=3.8cm]{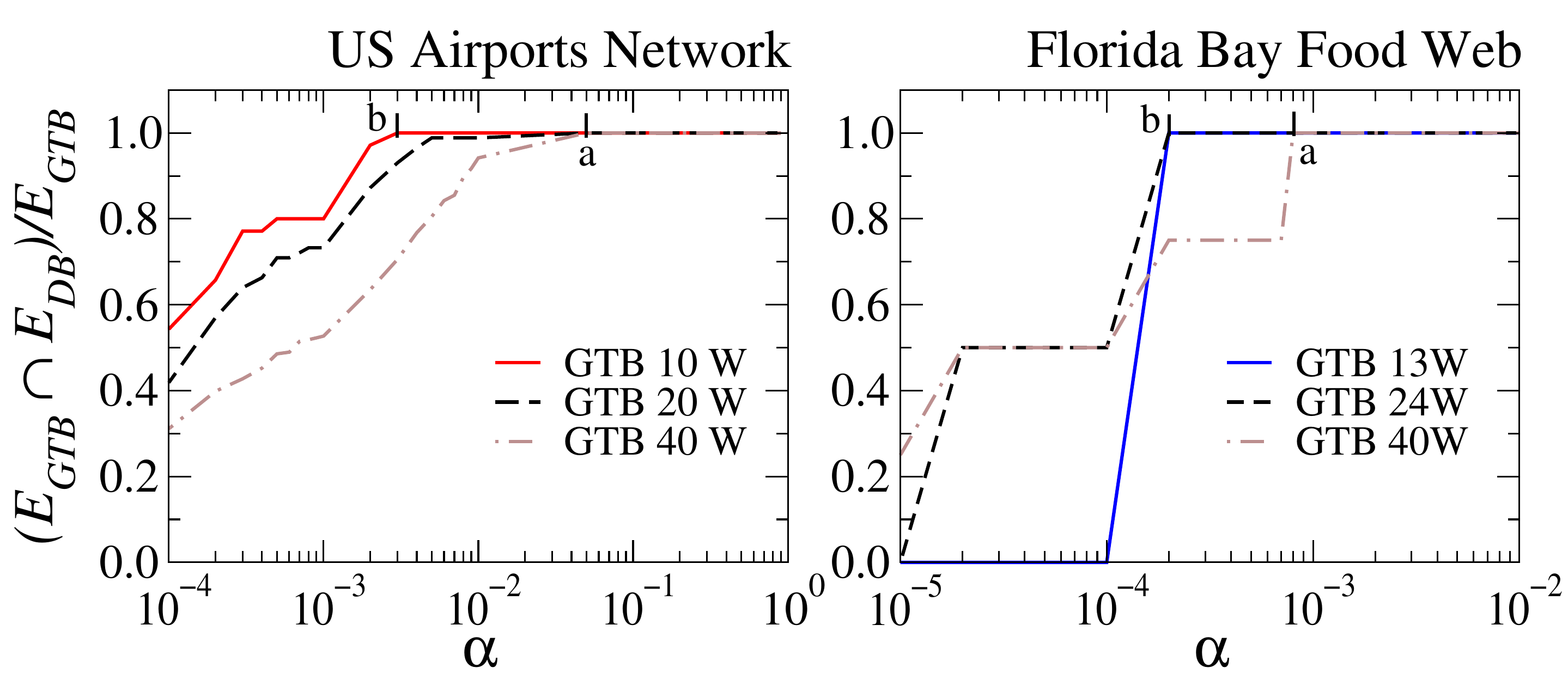}
\end{center}
\caption{Fraction of edges in different Global Threshold backbones
  (GTB) included in the Disparity backbone (DB) as a function of the
  significance level. As shown, points {\bf a} and {\bf b} in the US
  airport network mark Disparity backbones including a $100\%$ of the
  40-W and 10-W Global Threshold backbones, respectively; points {\bf
    a} and {\bf b} in the Florida Bay food web mark Disparity
  backbones including a $100\%$ of the 40-W and 13-W Global Threshold
  backbones, respectively. See also Table~1.}
\label{fig:3}
\end{figure}

The Florida Bay food web is a directed network (see Supplementary Information for
an explanation of the methodology in the case of weighted directed neworks). We
draw its multiscale backbone for $\alpha=0.0008$, which contains the top $40\%$ of
heaviest links (see entry (a) in Table~1 and Fig.~3). Notice that, in this case,
the concentration of weight in a few links is so important that the represented
disparity backbone contains approximately half of the total weight in the network.
Again, star motifs are uncovered, formed by mainly incoming connections -like for
the pelican- or mainly outgoing ones -bivalves. More in general, specific
subsystems dominated by significant fluxes can be easily identified, which might be
an evidence of a historical evolution of the network from smaller modular and
disconnected structures to the complete ecosystem we observe today. Another
interesting remark refers the presence in the backbone of species with relatively
few trophic links. Species with few connections are usually assumed to have a low
impact on the ecosystems. However, counterexamples can be found and such species
may act as the structural equivalent of keystone species, whereas species with many
trophic linkages may be more conceptually similar to dominant
species~\cite{Dunne:2002}. Due to its local approach, our filter mixes both types
in the backbones, where simultaneously coexist big hubs --like the Predatory
Shrimp, which in the complete network approximately has an average number of
incoming connections and the maximum number of outgoing ones, 13 and 61
respectively-- with more modest species in terms of connections --like Benthic
Flagellates, with in-degree 1 and out-degree 10, both below the average.

\subsection{Conclusions}
The disparity filter exploits local heterogeneity and local correlations among
weights to extract the network backbone by considering the relevant edges at all
the scales present in the system. The methodology preserves an edge whenever its
intensity is a statistically not compatible with respect to a null hypothesis of
uniform randomness for at least one of the two nodes the edge is incident to, which
ensures that small nodes in terms of strength are not neglected. As a result, the
disparity filter reduces the number of edges in the original network significantly
keeping, at the same time, almost all the weight and a large fraction of nodes. As
well, this filter preserves the cut-off of the degree distribution, the form of the
weight distribution, and the clustering coefficient.
\begin{figure*}[t]
\begin{center}
\includegraphics[height=5.5cm]{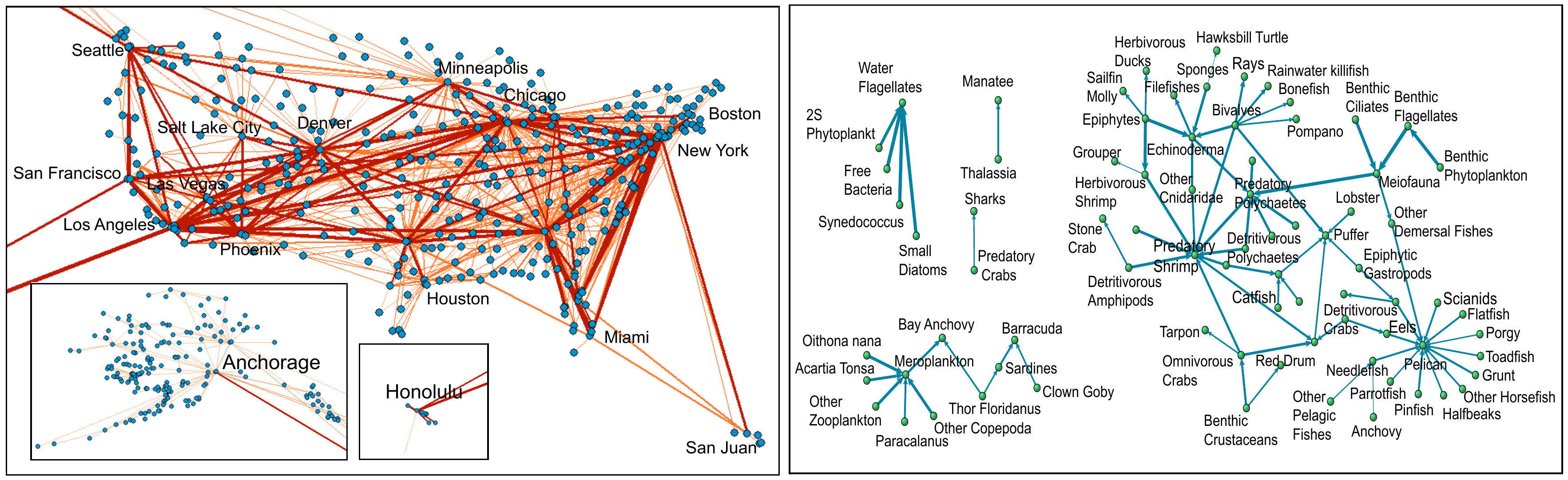}
\end{center}
\caption{Pajek representations~\cite{Pajek:2003} of disparity
  backbones. {\bf Top}. The $\alpha=0.003$ multiscale backbone of the
  2006 domestic segment of the U.S. airport transportation system. This
  disparity backbone includes entirely the top $10\%$ of the heaviest
  edges. {\bf Bottom}. The $\alpha=0.0008$ multiscale backbone of the
  Florida Bay ecosystem in the dry season. This disparity backbone
  includes entirely the top $40\%$ of the heaviest edges. These
  disparity backbones correspond to points (b) for the US airport
  network and (a) for the Florida Bay food web in Table~1 and
  Fig.~3. The connection with maximum weight for the US airport
  network is Atlanta-Orlando, with value $\omega_{max}=1,290,488$ $pasengers/year$ and for the Florida Bay Food Web Free Bacteria to
  Water Flagellates with value $\omega_{max}=12.90$ $mgCy^{-1}m^{-2}$.} \label{fig:4}
\end{figure*}

As a criticism, one could say that it only works in the case of systems with strong
disorder, where the weights are heterogeneously distributed both at the global and
local level. Nevertheless, all filters present limitations, one has to take them
into account in relation to the problem under analysis. Which strategy is the most
appropriate for a particular problem should be carefully judged and we cannot
exclude the possibility that a combination of different techniques turns out to be
the most appropriate. Yet, the ubiquitous presence of fluctuations and disorder
spanning many length scales uncovered in many real networks provides a wide range
of potential applications for the present methodology in biology (metabolic
networks, brain, periodically regulated genes), information technology (Internet,
World Wide Web), economics (World Trade Web) and finance (stocks markets).

\begin{materials}
\section{Local heterogeneity of edges' weight} In order to asses the
effect of inhomogeneities in the weights at the local level, for each node $i$ with
$k$ neighbors one can calculate the function~\cite{Guichard:2003,Almaas:2004}
\begin{equation}
\Upsilon_i(k) \equiv kY_i(k)= k\sum_{j}p_{ij}^2.
\label{eq:Y}
\end{equation}
The function $Y_i(k)$ has been extensively used in several fields as a standard
indicator of concentration for more than half a century: in
Ecology~\cite{Simpson:1949}, Economics~\cite{HHIHerfindahl:1959},
Physics~\cite{Derrida:1987} and recently in the Complex Networks literature where
it is known as the disparity measure~\cite{Guichard:2003}. In all cases, $Y_i(k)$
characterizes the level of local heterogeneity. Under perfect homogeneity, when all
the links share the same amount of the strength of the node, $\Upsilon_i(k)$ equals
1 independently of $k$, while in the case of perfect heterogeneity, when just one
of the links carries the whole strength of the node, this function is
$\Upsilon_i(k)=k$. An intermediate behavior is usually observed in real systems
with $\Upsilon_i(k)\propto k^\alpha$ and the exponent close to $1/2$. In this case,
the weights associated to a node are then peaked on a small number of links with
the remaining connections carrying just a small fraction of the node's strength.
This is the situation where our filter will be more useful, highlighting structures
impossible to detect using the global threshold filter. In this way, the disparity
function can be used as a preliminary indicator of the presence of local
heterogeneities.
\begin{figure}[h]
\vspace {-0.7cm}
\begin{center}
\includegraphics[height=8.5cm]{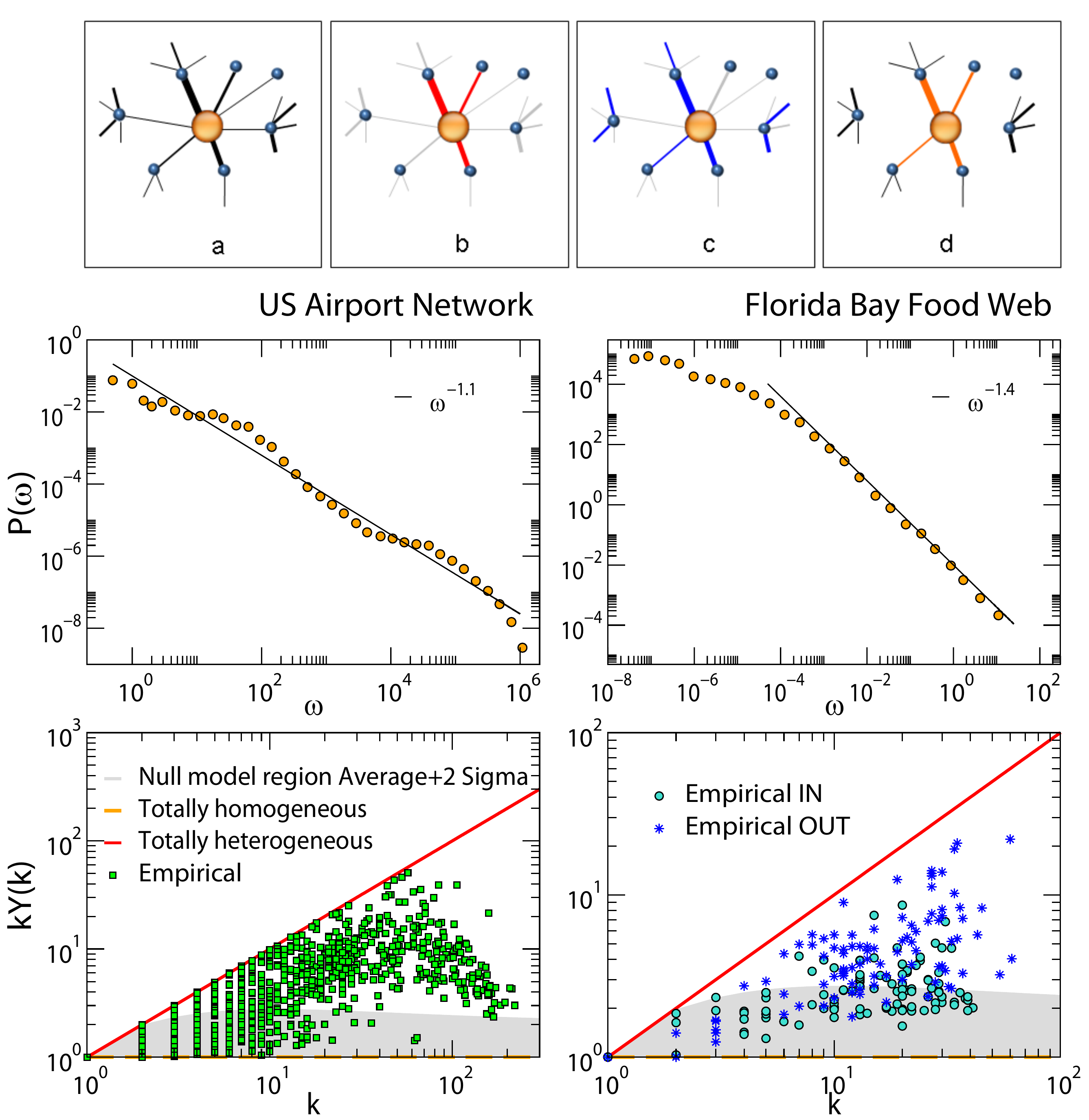}
\end{center}
\caption{{\bf Top sketch.} Sequential diagram illustrating the disparity filtering technique at the local level. We focus on the central node in orange and its first neighborhood. a) original network; b) edges of the central node with weights that are statistically significant heterogeneity; c) the same for the neighbors; d) intersection of the colored edges in b) an c) that are finally selected in the backbone. {\bf Middle graphs.} Distribution of link's weights spanning for six
  decades. Even though this distribution does not have a clear
  functional form, a direct power law fit of the form
  $\omega^{-\beta}$  yields an exponent $\beta=1.1$, so with a
  diverging first moment. {\bf Bottom graphs.} Scattered plot of the
  disparity measure for individuals airports of the US airport
  network. The grey area corresponds to the average plus 2 standard
  deviations given by the null model.} \label{panel1}
\end{figure}

\section{The null model} The probability density function of Eq.~(\ref{eq:PX}),
along with the join probability distribution for two intervals given by
\begin{equation}
\rho(x,y)dxdy=(k-1)(k-2)(1-x-y)^{k-3}\Theta(1-x-y)dxdy,
\label{eq:PXY}
\end{equation}
where $\Theta(\cdot)$ is the Heaviside step function, can be used to calculate the
statistics of $\Upsilon_{null}(k)$ for the null model. The average
$\mu(\Upsilon_{null}(k))= k \mu(Y_{null}(k))$ and the variance
$\sigma^2(\Upsilon_{null}(k))= k^2 \sigma^2(Y_{null}(k))$ are found to be:
\begin{eqnarray}
\mu(\Upsilon_{null}(k)) &=& \frac{2k}{k+1} \\
\sigma^2(\Upsilon_{null}(k))&=&
k^2 \left(\frac{20+4k}{(k+1)(k+2)(k+3)}-\frac{4}{(k+1)^2}\right).
\label{eq:NMY}
\end{eqnarray}
Notice that the two moments depend on the degree $k$ so that each node in the
network with a certain degree $k$ should be compared to the corresponding null
model.

The observed values $\Upsilon_{ob}(k)$ compatible with the null hypothesis could be
defined as those in the region between $\langle \Upsilon_{null}(k)\rangle+a \cdot
\sigma\left(\Upsilon_{null}(k)\right)$ and perfect homogeneity, so that local
heterogeneity will be recognized only if the observed values lie outside this area,
\begin{equation}
\Upsilon_{ob}(k)>\mu(\Upsilon_{null}(k))+a \cdot
\sigma\left(\Upsilon_{null}(k)\right).
\end{equation}
The variable $a$ is a constant determining the confidence interval for the
evaluation of the null hypothesis. The larger it is the more restrictive becomes
the null model and the more disordered weights should be for local heterogeneity to
be detected. A typical value in analogy to gaussian statistics could be for
instance $a=2$.

As shown in Fig.~5, the overall distributions of weights for both networks
considered here are very broad, with tails approaching power-law behaviors spanning
six decades for the U.S. airport network and more than four for the Florida Bay
food web. At the local level, $\Upsilon(k)$ measurements cannot be explained by the
null model for most nodes.
\end{materials}

\begin{acknowledgments}
M. A. S. acknowledges support by DGES grant No. FIS2007-66485-C02-01, M. B. by DGES
grant No. FIS2007-66485-C02-02. A.V. is partially supported by the NSF award
IIS-0513650 and NIH R21-DA024259.
\end{acknowledgments}

\newpage
\section{SUPPORTING INFORMATION}

\section{The disparity filter for directed weighted networks}

In many systems, interactions between pairs of elements are asymmetric, running
partial or totally in one of the two possible directions. Noticeable examples are
the World Wide Web~\cite{Serrano:2007a}, email networks~\cite{Newman:2002c},
citation networks~\cite{Redner:1998}, genetic and metabolic
networks~\cite{Jeong:2000,Almaas:2004}, or economic networks such as the World
Trade Web~\cite{Serrano:2007b}, among others. The undirected network representation
becomes then a first order approximation that can be refined by representing the
connections as arrows, indicating the source node at the tail and the destination
node at the head. In this way, directed network representations are more complete
and convey more information about the system when directionality of the
interactions is relevant. This increase of information content is reflected at the
simplest level even in the description of the nodes' connectivities, so that each
vertex has to be described by two coexisting degrees $k^{in}$ and $k^{out}$
representing the number of incoming neighbors pointing to it and the number of
outgoing neighbors pointed by it respectively, which sum up to the total degree
$k=k^{in}+k^{out}$. Hence, the degree distribution for a directed network is a
joint degree distribution $P(k^{in},k^{out})$ of in- and out-degrees, which in
general may be correlated. In the following, we assume they are not.

Our filtering methodology to extract the backbone of relevant connections in
complex multiscale networks can be extended to weighted directed networks. In this
type of representations, the total strength $s_i$ associated to a certain node $i$
has two contributions coming from the incoming strength $s_i^{in}$ and the outgoing
strength $s_i^{out}$, which are obtained by summing up all the weights of the
incoming or outgoing links respectively. The normalized weights of edges linking
node $i$ with its neighbors are calculated as $p_{ij}^{in} = w_{ij}^{in}/s_i^{in}$
if the link corresponds to an incoming connection, and $p_{ij}^{out} =
w_{ij}^{out}/s_i^{out}$ if it is associated to an outgoing one, being $w_{ij}^{in}$
the weight of the incoming connection to its neighbor $j$ and $w_{ij}^{out}$ the
weight of the outgoing one. Take into account that the incoming connection from the
point of view of the head node is at the same time an outgoing connection of the
tail node.

The strategy in this case is as before based on the detection local
heterogeneities. The goal is to preserve the edges carrying a weight that
represents a local significant deviation with respect to a statistical null model
for the local assignment of weights by using the disparity function. But this time
with the condition that incoming and outgoing links associated to a node must be
considered separately. For each node $i$ with $k^{in}$ incoming neighbors and
$k^{out}$ outgoing ones, one can calculate the functions
\begin{eqnarray}
\Upsilon_i(k^{in}) &\equiv& k^{in}Y_i(k^{in})= k^{in}\sum_{j}(p_{ij}^{in})^2,
\label{eq:Y2}\\
\Upsilon_i(k^{out}) &\equiv& k^{out}Y_i(k^{out})= k^{out}\sum_{j}(p_{ij}{out})^2.
\end{eqnarray}
$Y_i(k^{in})$ characterizes the level of local heterogeneity in the incoming
weights while $Y_i(k^{out})$ correspond to the outgoing counterpart. As happens in
the undirected case, under perfect homogeneity, when all the incoming (outgoing)
links share the same amount of the incoming (outgoing) strength of the node,
$\Upsilon_i(k^{in})$ ($\Upsilon_i(k^{out})$) equals 1 independently of $k^{in}$
($k^{out}$), while in the case of perfect heterogeneity, when just one of the
incoming (outgoing) links carries the whole incoming (outgoing) strength of the
node, this function is equal to $k^{in}$ ($k^{out}$). An intermediate power law
behavior is usually observed in real systems indicating that the incoming
(outgoing) weights associated to a node are peaked on a small number of links with
the remaining connections carrying just a small fraction of the node's incoming
(outgoing) strength. This is the situation where our filter will be more useful,
highlighting structures impossible to detect using the global threshold filter. In
this way, the disparity function can be used as a preliminary indicator of the
presence of local heterogeneities.

\subsection{The null model}
The null model that we use to define anomalous fluctuations of weights in directed
networks with strong disorder provides the expectation for the disparity measures
above in a pure random case. The null hypothesis is made independently for the set
of incoming and outgoing connections and is the same as in the undirected case. It
assumes that the normalized weights which correspond to the incoming (outgoing)
connections of a certain node of in-degree $k^{in}$ ($k^{out}$) are produced by a
uniform random assignment. To visualize this process, $k^{in}-1$ ($k^{out}-1$)
points are distributed with uniform probability in the interval $[0,1]$ so that it
ends up divided in $k^{in}$ ($k^{out}$) subintervals. Their lengths would represent
the expected values for the $k^{in}$ ($k^{out}$) normalized weights $p_{ij}^{in}$
($p_{ij}^{out}$) according to the null hypothesis. The incoming and outgoing
probability density functions for one of these variables taking a particular value
$x$ is
\begin{equation}
\rho(x)dx=(\kappa-1)(1-x)^{\kappa-2}dx, \label{eq:PX2}
\end{equation}
where $\kappa$ stands for $k^{in}$ or $k^{out}$ as the fluctuations in incoming or
outgoing intensities are being evaluated. This probability density function, along
with the join probability distribution for two intervals given by
\begin{equation}
\rho(x,y)dxdy=(\kappa-1)(\kappa-2)(1-x-y)^{\kappa-3}\Theta(1-x-y)dxdy,
\label{eq:PXY2}
\end{equation}
where $\Theta(\cdot)$ is the Heaviside step function, can be used to calculate the
statistics of $\Upsilon_{null}(k^{in})$ and $\Upsilon_{null}(k^{out})$ for the null
model. The averages $\mu(\Upsilon_{null}(\kappa))= \kappa \mu(Y_{null}(\kappa))$
and the standard deviations $\sigma^2(\Upsilon_{null}(\kappa))= \kappa^2
\sigma^2(Y_{null}(\kappa))$ are found to be:
\begin{eqnarray}
\mu(\Upsilon_{null}(\kappa)) =& \hspace{-3.9cm} \frac{2\kappa}{\kappa+1} \\
\sigma^2(\Upsilon_{null}(\kappa))=&
\kappa^2 \left(\frac{20+4\kappa}{(\kappa+1)(\kappa+2)(\kappa+3)}-\frac{4}{(\kappa+1)^2}\right).
\label{eq:NMY2}
\end{eqnarray}
Notice that the two moments depend on the incoming or outgoing degree $\kappa$ so
that each node in the network with a certain $k^{in}$ and $k^{out}$ should be
compared to the corresponding functions.

In real or modeled networks, the disparities can be directly observed and the
functions $\Upsilon_{ob}(k^{in})$ and $\Upsilon_{ob}(k^{out})$ can be compared
against the null model expectations. Values compatible with the null hypotheses
could be defined as those in the region between $\langle
\Upsilon_{null}(\kappa)\rangle+a \cdot \sigma\left(\Upsilon_{null}(\kappa)\right)$
and perfect homogeneity, so that local heterogeneity will be recognized only if the
observed values lie outside this area,
\begin{equation}
\Upsilon_{ob}(\kappa)>\mu(\Upsilon_{null}(\kappa))+a \cdot
\sigma\left(\Upsilon_{null}(\kappa)\right).
\end{equation}
The parameter $a$ is a constant determining the confidence interval for the
evaluation of the null hypothesis. The larger it is the more restrictive becomes
the null model and the more disordered weights should be for local heterogeneity to
be detected. A typical value in analogy to gaussian statistics could be for
instance $a=2$. In this way, it is possible to characterize quantitatively the
level of disorder observed in the distribution of weights in incoming and outgoing
links. Specially when this disorder is high, our disparity filtering technique
allows us to extract the backbone of relevant directed connections.

\subsection{The disparity filter} The disparity filter proceeds by identifying which incoming and outgoing links for each node
should be preserved in the network. The null model allows this discrimination by
the calculation for each incoming (outgoing) edge of a a given node $i$ of the
corresponding probability $\alpha_{ij}^{in}$ ($\alpha_{ij}^{out}$) that its
normalized weight $p_{ij}^{in}$ ($p_{ij}^{out}$) is compatible with the null
hypothesis. In statistical inference, this concept is known as the $p$-value, the
probability that if the null hypothesis is true one obtains an o value for the
variable under consideration larger or equal than the observed one. By imposing a
significance level $\alpha$, the incoming (outgoing) links that carry weights which
can be considered not compatible with a random distribution can be filtered out
with an certain statistical significance. All the incoming (outgoing) links with
$\alpha_{ij}^{in}<\alpha$ ($\alpha_{ij}^{out}<\alpha$) reject the null hypothesis
and can be considered as significant heterogeneities. By changing the significance
level we can filter out the incoming (outgoing) links progressively focusing on
more relevant heterogeneities. Statistically significant inhomogeneous weights will
be then those which satisfy
\begin{eqnarray}
\alpha_{ij}^{in}=1-(k^{in}-1)\int_0^{p_{ij}^{in}} (1-x)^{k^{in}-2}dx < \alpha,\\
\alpha_{ij}^{out}=1-(k^{out}-1)\int_0^{p_{ij}^{out}} (1-x)^{k^{out}-2}dx < \alpha.
\label{eq:confidencelevel2}
\end{eqnarray}
Note that these expressions are calculated as a function of the probability density
function Eq.~[\ref{eq:PX2}], and again depend on the number of connections $k^{in}$
or $k^{out}$ of the node to which the directed link under consideration is
attached.

The multi-scale backbone of weighted directed networks is then obtained by
preserving all the incoming and outgoing links which beat the threshold for at
least one of the two nodes at the ends of the link while discounting the rest.
Notice that an outgoing connection for the tail node is an incoming connection for
the head one, so the outgoing connections and the appropriate null model should be
considered for the first while incoming connections and the corresponding null
model for the second. In the case of a node $i$ with out-degree $k_{i}^{out}=1$
connected to a node $j$ with in-degree $k_j^{in}>1$, we keep the connection only if
it beats the threshold for the in-null model of node $j$, while if the in-degree of
node $i$ $k_{i}^{in}=1$ and it is connected to a node $j$ of out-degree
$k_j^{out}>1$, we keep the connection only if it beats the threshold for the
out-null model of node $j$. In this way, relevant fluctuations at all scales are
selected and small nodes in terms of strength are not belittled so that the system
remains in the percolated phase. Finally, in the rare case than node $i$ has
out-degree $k_{i}^{out}=1$ and in-degree $k_i^{in}>1$ and is connected to a node
$j$ with in-degree $k_j^{in}=1$ and out-degree $k_j^{out}>1$, we keep the
connection as it is the only way to maintain the connectivity of the network.

By choosing a constant significance level $\alpha$ we obtain a homogeneous criteria
that allows us to compare inhomogeneities in nodes with different magnitude in
connections and strength. Decreasing the statistical confidence more restrictive
subsets are obtained, giving place to a potential hierarchy of backbones. This
strategy will be efficient whenever the level of heterogeneity is high. Otherwise,
the pruning could lose its hierarchical attribute.

\section{Networks with uncorrelated weights}
The disparity filter and the global threshold strategy give similar results when
applied to a complex network with uncorrelated weights, whenever their probability
distribution $P(\omega)$ has a well defined average. From a practical point of
view, a network with uncorrelated weights can be easily realized by assigning to
each edge of the network an intensity drawn independently at random from
$P(\omega)$. Distributions with a well defined average could be homogeneous
distribution, where all weights fluctuate around a characteristic value, but could
also be highly heterogeneous ones, for instance those with power-law form with
exponent larger than two.

Next, we prove analytically --for undirected networks although the same reasoning
is also valid for directed ones-- the approximate equivalence of the two models for
a certain relation between the significance level $\alpha$ and the global threshold
$\omega_c$, that we derive. More specifically, we demonstrate that the probability
for a given edge of weight $\omega_{ij}$ connected to a node $i$ of degree $k$ of
remaining in the disparity-filtered network $S(\omega_{ij}|k)$ is the same as that
of remaining in the globally thresholded one $\Theta(\omega_{ij}-\omega_c)$, where
$\Theta(\cdot)$ is the Heaviside step function. Henceforth, we generally refer to
these probabilities as survival probabilities.

From Eq.[2] in the main text, the disparity filter keeps those edges with weights
$\omega_{ij}>(\alpha^{-1/(k-1)}-1)\sum_{l \neq j}\omega_{il}$. The disparity filter
survival probability can thus be expressed as
\begin{eqnarray}
S(\omega_{ij}|k)=\int\cdots\int \Theta\left(\omega_{ij}-(\alpha^{-1/(k-1)}-1)\sum_{l\neq j}\omega_{il}\right) \cdot \nonumber \\ \cdot \prod_{l \neq j}P(\omega_{il})d\omega_{il}.
\label{survival}
\end{eqnarray}
In the previous equation, we have taken into account that, for this particular
model, weights are uncorrelated and that for every edge the weight is identically
and independently distributed according to $P(\omega)$. Calculations are very much
simplified in the Laplace space where, generically, we define the Laplace transform
of a function $f(\omega)$ as $\hat{f}(u)=\int_0^{\infty} f(\omega) e^{-u \omega}d
\omega$. Using this transformation, equation (\ref{survival}) reads
\begin{equation}
\hat{S}(u|k)=\frac{1}{u} \hat{P}\left[ u (\alpha^{-1/(k-1)}-1)\right]^{k-1}.
\label{eq:psk}
\end{equation}
For large degrees, one can make the approximation
\begin{equation}
\hat{P}\left[u (\alpha^{-1/(k-1)}-1)\right] \simeq  \hat{P}\left[ u \ln{\alpha^{-1}}/(k-1)\right],
\end{equation}
and truncating the Taylor series expansion to the first order in $u$
\begin{equation}
\hat{P}\left[u (\alpha^{-1/(k-1)}-1)\right] \simeq 1-\langle \omega \rangle u\ln{\alpha^{-1}}/(k-1).
\end{equation}
Substituting this into Eq.~[\ref{eq:psk}],
\begin{eqnarray}
\hat{S}(u|k)&\simeq&\frac{1}{u}\left[ 1 - \langle \omega \rangle \frac{u\ln{\alpha^{-1}}}{(k-1)}\right]^{k-1} \nonumber \\
&\simeq&\frac{1}{u}e^{- \langle \omega \rangle u\ln{\alpha^{-1}}}.
\end{eqnarray}
Notice that this expression has lost any dependence on the vertex degree $k$.
Finally, inverting the Laplace transformation
\begin{equation}
S(\omega_{ij}|k) \simeq \Theta(\omega_{ij}-\langle \omega \rangle \ln{\alpha^{-1}}).
\end{equation}
\begin{figure}[h]
\begin{center}
\includegraphics[height=6.1cm]{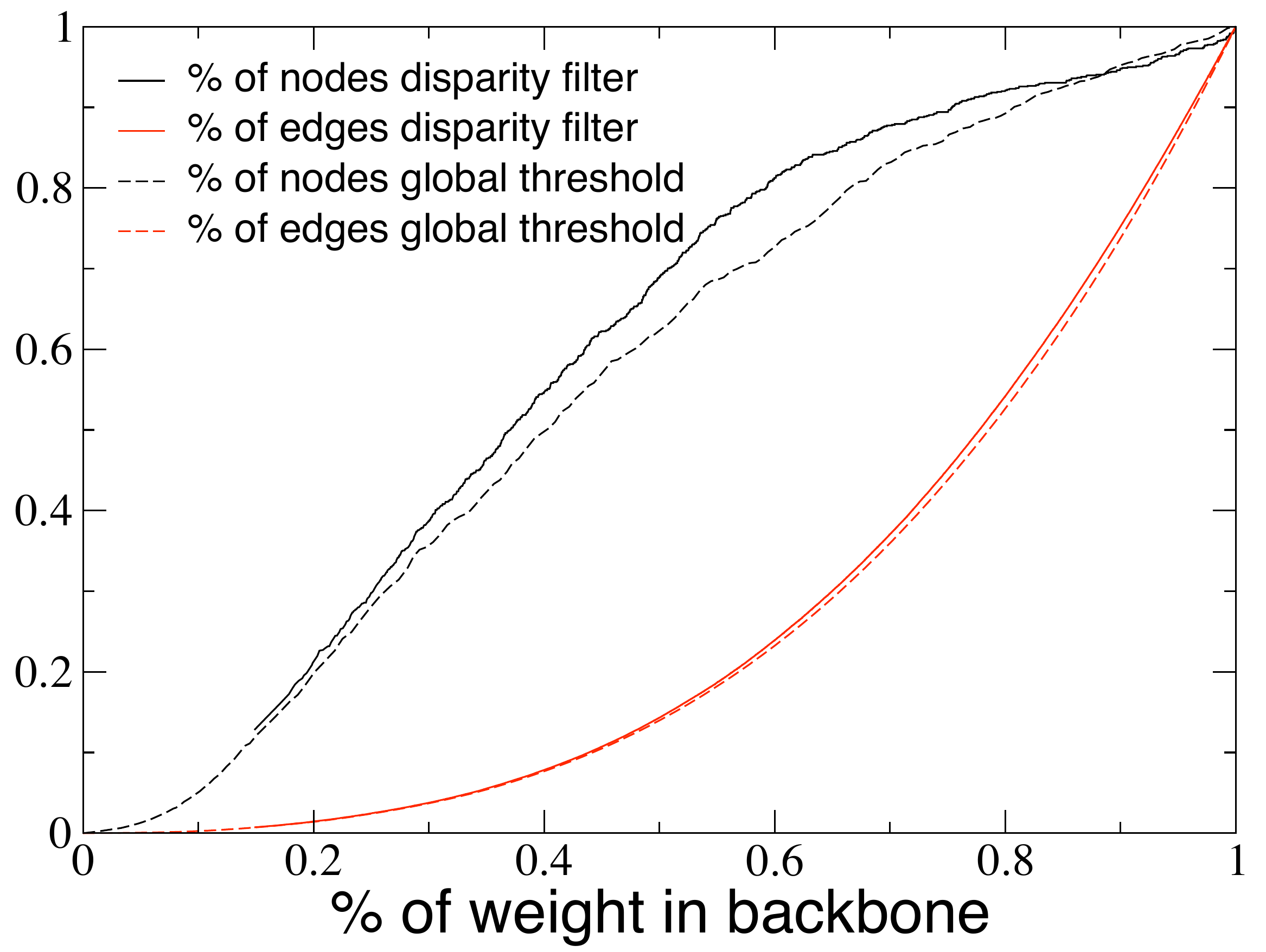}
\end{center}
\caption{Fraction of nodes and edges as a function of the fraction of total weight retained by the global and disparity filters acting on the airport network with a random assignment of weights according to the distribution $P(\omega) \propto \omega^{-2.5}$.}
\label{fig:1si}
\end{figure}
Hence, the survival probability under the disparity filter with significance level
$\alpha$ is approximately equal to the survival probability under the global
threshold for a threshold value $\omega_c=\langle \omega \rangle \ln{\alpha^{-1}}$,
independent of the degree $k$. Figure~1 shows the result of both filters on the
airport network with a random assignment of weights to edges. In this case, we use
$P(\omega) \propto \omega^{-\beta}$ with $\beta=2.5$. As it is clearly visible,
both filters give very similar results, in agreement with the calculations above.

\subsection{Unbounded average}
Notice that if the average of $P(\omega)$ is unbounded, the previous relation is
not well defined. However, this is the case of most real networks, that are
characterized by a weight distribution that is power-law with exponent less than
two, so that its first moment diverges. In this situation, the equivalence of the
two methodologies does not hold. This is mainly due to the symmetry breaking that
we impose on the filtering condition when we consider that the same intensity $w$
may be relevant in a different way if considered as associated to $\omega_{ij}$ and
$\omega_{ji}$. Each edge is incident to two nodes; while the weight carried by the
edge may not be a relevant fluctuation for one node (for instance a node with
several other links with large weight) it could be a relevant fluctuation for the
other node. This is what allows us to preserve relevant fluctuations at different
scales and providing a backbone including nodes handling a total weight of very
different magnitude.
\begin{figure}[t]
\begin{center}
\includegraphics[height=6.1cm]{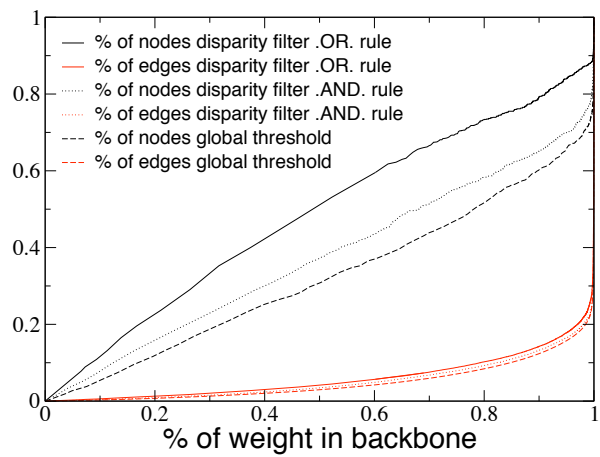}
\end{center}
\caption{Fraction of nodes and edges as a function of the fraction of total weight retained by the global and disparity filters (with .OR. and .AND. rules) acting on the airport network with reshuffled weights.}
\label{fig:2si}
\end{figure}

For this reason, instead of considering weights directly, our methodology works
with the normalized weights $p_{ij}=\omega_{ij}/s_i$ and $p_{ji}=\omega_{ij}/s_j$
as independent quantities. One might want to enforce symmetry by imposing a rule
{\it AND} instead of the rule {\it OR} that we have chosen, so that a connection is
preserved whenever its intensity is significant for both nodes involved. However,
the rule {\it OR} in the disparity filter, that we prefer because it ensures that
small nodes in terms of strength are not belittle, only demands that the connection
is important for one of the two. Remember that in networks where weights are not
correlated there is a relation between the strength $s$ of nodes and the average
weight in the network of the form $s\simeq k\langle \omega \rangle$. If the average
is not well defined, the strength of nodes can fluctuate wildly so that the same
weight can be experienced as extremely important or unimportant depending on the
node and, as a consequence, the rules {\it AND} and {\it OR} produce very different
results.

In Fig.~2, we show the effect of considering the disparity filter with rules {\it
AND} and {\it OR} on networks with uncorrelated weights with unbounded average. The
{\it AND} disparity filter is qualitatively very similar to the global threshold
algorithm regarding number of preserved nodes and edges, while the {\it OR}
disparity filter maintains a similar number of edges with a much larger number of
nodes.

\end{article}

\end{document}